\pdfoutput=1
\documentclass[preprint,11pt,3p,nopreprintline]{elsarticle}

\usepackage{graphicx}
\usepackage{tabularx}
\usepackage{amssymb}
\usepackage{paralist}
\usepackage{amsmath,mathtools}
\usepackage{makecell}
\usepackage{enumitem}
\usepackage{float}
\usepackage{hyperref}
\usepackage{multirow}
\usepackage[table]{xcolor}
\usepackage{pifont}          \usepackage{tikz}
\usetikzlibrary{shapes.geometric, arrows.meta, positioning, calc, fit, backgrounds}
\usepackage{csquotes}
\usepackage{rotating}
\usepackage{booktabs}        \usepackage{array}           \usepackage{longtable}
\usepackage{url}
\usepackage{doi}
\usepackage{subcaption}
\usepackage{bbm}
\usepackage{threeparttable}
\usepackage{nomencl}
\usepackage{arydshln}
\usepackage{algorithmicx,algorithm,algpseudocode}
\usepackage{siunitx}
\usepackage{lineno}          \modulolinenumbers[5]

\biboptions{sort&compress}
\definecolor{mygreen}{rgb}{0,0.6,0}

\algdef{SE}[DOWHILE]{Do}{doWhile}{\algorithmicdo}[1]{\algorithmicwhile\ #1}\algrenewcommand\algorithmicrequire{\textbf{Input:}}
\algrenewcommand\algorithmicensure{\textbf{Output:}}

\makenomenclature

\allowdisplaybreaks

\hypersetup{
    colorlinks = true,
    urlcolor   = blue,
    linkcolor  = blue,
    citecolor  = blue
}

\begin{document}

\begin{frontmatter}

\title{Impact of the Performance of Autonomy and Communication Subsystems on the Level-of-Safety of Partially Autonomous Machine Systems}
\author[inst1]{Yutian Pang\corref{cor1}}
\cortext[cor1]{Corresponding author.}
\ead{yutian.pang@austin.utexas.edu}
\author[inst1]{John-Paul Clarke}

\affiliation[inst1]{organization={Department of Aerospace Engineering and Engineering Mechanics},
            addressline={The University of Texas at Austin},
            city={Austin},
            postcode={78712},
            state={TX},
            country={USA}}

\begin{highlights}
\item An analytical model gives the level-of-safety of partially autonomous systems.
\item Mean time between interventions (MTBI), urgency, and complexity describe autonomy.
\item A dimensionless criticality ratio prescribes the shape of the safety tradeoff.
\item A perfect link cuts the required MTBI a hundredfold at a ratio of 0.01, half at 1.
\item Risk maps separate compliance in expectation from compliance with assurance.
\end{highlights}

\begin{abstract}
The level-of-safety of a partially autonomous machine system depends on two subsystems: the autonomy subsystem that manages the system between human interventions; and the communication subsystem that enables those interventions. These subsystems are typically evaluated independently. We model the level-of-safety analytically as a function of their joint performance. Three parameters characterize the autonomy subsystem: the Mean Time Between Interventions (MTBI); the urgency $\mathcal{U}$ or mean time within which an intervention must be completed to avoid failure; and the complexity $\mathcal{C}$ or mean time to understand, decide on and execute it. Two characterize the communication subsystem: its availability $P_A$ and its latency $\varepsilon$. A dimensionless criticality ratio $\rho=(\mathcal{C}+\varepsilon)/\mathcal{U}$ prescribes the shape of the Pareto tradeoff between the MTBI and $P_A$ at a constant level-of-safety. At $\rho=0.01$, a perfect link lowers the required MTBI by two orders of magnitude. At $\rho=1$, it only halves it. Under exponential timing assumptions the model is in closed form, and compliance with a safety budget can be stated at a chosen assurance level. We illustrate it for remotely piloted aircraft with assumed parameter values and propose a simulation procedure to estimate them.
\end{abstract}


\end{frontmatter}

\section{Introduction \label{sec: introduction}}
In a growing number of systems, the human operator no longer sits \textit{in the control loop}. Rather, the automation (or autonomy subsystem) manages the system on its own and only requests help from the operator when it finds itself in a situation that it cannot handle. Within the aviation domain, the Federal Aviation Administration (FAA) describes this arrangement as a \textit{human-over-the-loop} model, where human oversight is triggered only to handle exceptions \citep{faa2023uamconops}. To this end, the National Aeronautics and Space Administration (NASA) is exploring the use of remote, ground-based pilots for uncrewed cargo aircraft \citep{hayashi2022paav}. NASA also envisions a single individual overseeing several aircraft \citep{goodrich2021uml}. This vision is consistent with the proposed FAA rule for uncrewed aerial system (UAS) operations beyond visual line of sight where direct manual control will not be allowed once the aircraft is beyond visual line of sight, and flight coordinators will (akin to an air traffic controller) monitor all aircraft in their jurisdiction and intervene if necessary \citep{faa2025bvlos}.
The same arrangement appears outside aviation. When automated road vehicles find themselves in situations they cannot resolve, they revert to a pool of remote supervisors \citep{hickert2023cooperation}. Autonomous ships depend on an operator whose response must arrive before the automation runs out of time \citep{rodseth2022towards}. Crews on missions beyond low Earth orbit face anomalies with communication delays of several minutes to mission control \citep{frank2013autonomous,valinia2022safe}.

In these partially autonomous machine systems, safety depends on the autonomy subsystem, which manages the system between human interventions; the communication subsystem, which carries the requests for help to the operator and the corrective commands back to the vehicle; and whether corrective actions can be completed before the situation is irresolvable. \citet{lachter2021box} refer to the set of situations that the automation can handle alone as its competency box. When a situation falls outside this set, authority passes to the human \citep{alves2018considerations}. Once that occurs, a narrow window remains before the situation can no longer be saved. We refer to this time window as the intervention window and to its mean length as the intervention urgency. Within that window, the operator must gain an understanding of the situation, decide on the appropriate action, and communicate the necessary commands to the vehicle for execution by the automation. We refer to the total time required to gain an understanding of the situation, decide on the appropriate corrective action, and execute the commands that are transmitted to the vehicle as the intervention complexity. And assuming the link is available, the time to communicate the commands to the vehicle is dependent on the latency of the communication link between the operator and the vehicle. Of course, the intervention will only succeed if the necessary commands are received and executed before the window closes.

The autonomy and communication subsystems share the responsibility for intervention. The autonomy subsystem determines how often an intervention is needed, how urgently the intervention must be made, and how long it takes to determine and execute the appropriate course of action. The communication subsystem determines how likely it is that there is a communication link available, and how much of a time delay is introduced by the link. A better autonomy subsystem asks for help less often; gives the operator more time for understanding, decision-making and execution; and makes the understanding, decision-making and execution easier. A better communication subsystem is available more often and has less latency. So, to some extent one can compensate for the other. This insight is consistent with the NASA maturity scale for Urban Air Mobility which warns that remote supervisors cannot be relied on to mitigate time critical safety concerns \citep{goodrich2021uml}. However, the NASA scale offers no criterion for deciding on these concerns. The answer depends on the vehicle, the function, and the link together.

In current practice, these two subsystems are not jointly evaluated. Link requirements come from the Required Communication Performance concept of the International Civil Aviation Organization \citep{icao2017pbcs}. It specifies a transaction time, a continuity, an availability, and an integrity for each type of transaction. The Joint Authorities for Rulemaking on Unmanned Systems (JARUS) carried these four parameters into the Required C2 Performance concept for the command and control (C2) link \citep{jarus2016rlp}. That document notes that the level of automatic operation of the aircraft drives the availability the link must reach. It then leaves each designer to settle the balance without a method. System safety assessment, meanwhile, assumes that the human intervention will succeed. The FAA guidance for Part 23 airplanes allows a probability of one for crew tasks judged reasonable \citep{faa2011ac231309}. The JARUS guidance for remotely piloted aircraft makes the same presumption for the remote pilot after a failure \citep{jarus2015amc1309}. That assumption is hard to justify when the person who must act is reached through a link with finite availability and delay. The proposed FAA rule acknowledges as much. It asks operators to assess how latency and intermittent lost link affect safety, and it leaves the metrics to an industry standard that has yet to be written \citep{faa2025bvlos}.

The autonomy subsystem is even harder to bring into such an assessment, because its performance is usually stated as an ordinal level \citep{parasuraman2000model,sae2021j3016,huang2005alfus}. An ordinal label says nothing about how often the system will need help, how urgently the help must be completed, and how long it takes to deliver the help. An ordinal label cannot for example be compared with a failure rate per flight hour. The National Research Council observed that the community has not developed quantitative levels of system reliability, nor a way to allocate such requirements among hardware, software, and pilots \citep{nrc2014autonomy}. Measures that count interventions are closer to what a safety assessment needs. Robotics counts operator interventions per unit time \citep{steinfeld2006common}, and automated driving reports the MTBI \citep{paz2020autonomous}. NASA's work on multi-vehicle operations uses the same quantities under the names mean neglect time and service time \citep{scheff2024mn}. So far these measures have been used to rank systems or to size operator teams, and none has been set against a safety budget.

The quantitative evidence on the coupling between autonomy and communication that exists is based on simulation, with the control architecture held fixed. Our earlier study modeled the availability of the C2 link as a two state Markov chain with transport latency \citep{pang2026reliability}. It then measured the mission success of a remotely piloted aircraft across a grid of availability and latency values. \citet{bulusu2022impact} ran a similar simulation for remotely piloted cargo aircraft. They observed losses of separation once the response time exceeded 30 s or the message drop probability exceeded 0.2. Both studies show how performance degrades as the link degrades. Neither can say how much onboard autonomy would be needed to make up for a poorer link, because the autonomy never changes. Some analytical studies do place autonomy and latency in one model \citep{safwat2025delays,tabassum2022dynamic}. They leave out the availability of the link, the rate at which help is needed, and a budget to compare the result against. To the authors' knowledge, no model maps the performance of the two subsystems to a level-of-safety that can be set against a safety budget. This paper asks two questions. \textit{What level-of-safety does a partially autonomous machine system reach for a given performance of its autonomy and communication subsystems? By how much can a better link relax the autonomy needed to meet a target level-of-safety?}

We answer both questions with an analytical model of the level-of-safety that has a closed form. The autonomy subsystem is described by three parameters, the MTBI, the urgency, and the complexity. The logarithm of the MTBI gives a continuous scale of intervention frequency. It covers the first of the three parameters, so designs with the same MTBI can still differ in urgency or complexity. The communication subsystem is described by two parameters, the availability and the latency of the link. The timing condition, that the time to understand, decide on, communicate and execute the intervention must fit within the intervention window, is the general time reliability model of probabilistic risk assessment (PRA) \citep{nasa2011pra}. When the times are exponentially distributed, the probability that the response fits has a closed form. The availability then scales that probability, and the rate at which interventions are needed turns the remaining probability of failure into a residual failure rate per hour of operation. That rate is the level-of-safety of the operation, and the safety budget is the target it has to meet.

The joint capabilities of the autonomy and communication subsystems relative to the timing condition are represented by the dimensionless criticality ratio $\rho$ that prescribes the shape of the Pareto tradeoff between the MTBI and the communication availability at a constant level-of-safety. The ratio is the mean time needed to understand, decide on, communicate and execute an intervention (the complexity plus the latency), divided by the urgency. When the ratio is small, a link with high availability lowers the MTBI needed to meet the prescribed level-of-safety. For example, at a ratio of 0.01, the MTBI of the autonomy subsystem can be two orders of magnitude less with a perfect communication subsystem than with no communication at all. As the ratio approaches one, remote intervention helps less and less. At a ratio of one even a perfect link only halves the required MTBI, so most of the safety budget has to be met by the autonomy subsystem. Because the autonomy parameters are uncertain, we propagate that uncertainty by sampling, which allows compliance with a budget to be stated at a chosen assurance level. We apply the model to remotely piloted aircraft with assumed parameter values and the aviation budgets per flight hour. We also describe a procedure that estimates the autonomy parameters of a given design from a closed loop simulator.

The contributions made in this paper are as follows.
\begin{itemize}
    \item We characterize the performance of the autonomy subsystem by three measurable parameters, the MTBI, the urgency, and the complexity. The MTBI is placed on a continuous scale through $\alpha=\log_{10}\mathrm{MTBI}$, with the MTBI in hours of operation, which are flight hours for aircraft. The scale expresses the frequency of interventions over orders of magnitude, and the three parameters can be estimated from simulated or operational data.
    \item We develop an analytical model of the level-of-safety of partially autonomous machine systems, under exponential timing assumptions. It expresses the level-of-safety as a residual failure rate that couples the MTBI, the urgency, and the complexity of interventions with the availability and latency of the communication link.
    \item We introduce the criticality ratio, and show that it prescribes the shape of the tradeoff between the MTBI and the availability at a constant level-of-safety. The constant risk frontiers that follow quantify how much a better link can lower the required MTBI for a given budget. The relief reaches two orders of magnitude at a ratio of 0.01 and falls to a factor of two at a ratio of one.
    \item We propagate the uncertainty in the autonomy parameters through the model to construct risk maps over the autonomy and communication parameters. The maps identify allocations that meet the safety budgets of the FAA and the European Union Aviation Safety Agency (EASA) in expectation or at a chosen assurance level.
    \item We introduce a closed loop simulation procedure that estimates the MTBI, the urgency, and the complexity for a given vehicle and function with an idealized link. The estimates separate the autonomy subsystem from the communication subsystem and replace the assumed values.
\end{itemize}

The remainder of the paper describes these contributions in five sections. Section~\ref{sec: literature review} reviews the related literature and states the research gap. Section~\ref{sec: methodologies} develops the model of the level-of-safety, the safety budget evaluated by sampling, and the constant risk frontier. Section~\ref{sec: results} gives a numerical illustration with assumed parameter values, and Section~\ref{sec: evaluation} describes the evaluation procedure that replaces those values. Section~\ref{sec: conclusion} concludes the paper and states its limitations.

\section{Literature Review  \label{sec: literature review}}
A model of the level-of-safety has to draw on literature that addresses the autonomy subsystem, human intervention, and the communication subsystem separately. We review these three areas in turn before examining the risk models that can combine them. The final subsection states the gap that the present model fills.

\subsection{Functional Authority and Measurable Autonomy}
Autonomy classifications provide a starting point by describing the authority assigned to humans and machines within a task. The scales of \citet{sheridan1978human} and \citet{parasuraman2000model} distinguish responsibility for selecting and carrying out actions. Later taxonomies apply related distinctions to driving, unmanned systems, and aviation \citep{vagia2016literature,sae2021j3016,huang2005alfus,anderson2018levels}. EASA likewise distinguishes artificial intelligence applications through their relationship with human authority \citep{easa2023airoadmap}. These descriptions help specify an operating architecture. \citet{endsley2017loa} further shows their value for understanding operator engagement, situation awareness, and effective oversight.

Applying these classifications across functions exposes the limits of a single ordinal description of autonomy. Critiques of autonomy levels emphasize this limitation and the difficulty of applying a single scale during design \citep{dsb2012autonomy,bradshaw2013seven,durst2014levels}. The National Research Council therefore uses the term increasingly autonomous without imposing a sharp boundary between automation and autonomy \citep{nrc2014autonomy}. NASA's autonomy taxonomy also defines independence in achieving goals without assigning universal levels \citep{fong2018taxonomy}. Function specific design and the multidimensional framework of \citet{idris2020autonomy} address this dependence on task and context \citep{proud2003methods}. A quantitative intervention measure must retain that scope.

Measuring intervention dependence within a specified function provides a quantitative complement to these contextual descriptions. Human and robot interaction studies use neglect time and interaction time to estimate how many robots an operator can support \citep{olsen2003metrics,crandall2005validating}. Extensions include operator waiting time in the supervision of multiple aircraft \citep{cummings2008predicting}. The intervention counts, the MTBI, and the neglect and service times named in the Introduction belong to the same family of measures \citep{steinfeld2006common,paz2020autonomous,scheff2024mn}. A reduced need for external intervention is also consistent with the control perspective on autonomy \citep{antsaklis2020autonomy}.

Using these intervention measures in a safety calculation also requires the urgency and complexity of the interventions and the performance of the communication subsystem. The same intervention frequency can produce a different level-of-safety when the urgency or the communication link differs. We use the logarithm of the MTBI to express intervention frequency over several orders of magnitude. The quantity applies to a specified function, alerting rule, and operating setting. The urgency and the complexity remain separate inputs to the model.

\subsection{Remote Intervention and Its Timing Limits}
For remotely piloted aircraft, the urgency and complexity of interventions depend on onboard competence and on the situations assigned to the human \citep{lachter2021box}. NASA's research on multiple aircraft identifies unresolved conflicts, delayed handoffs, and lost links as barriers to scaling remote supervision \citep{aubuchon2022mn,lacher2025mn}. Its cargo concept and tabletop exercises distinguish terminal operations from less urgent en route supervision \citep{hayashi2022paav,wolter2023paav}. \citet{sakakeeny2022framework} make functional allocation depend on time criticality, the operator to aircraft ratio, and loss of the C2 link. Their framework identifies the relevant dependencies qualitatively and leaves function complexity for further work.

The feasibility of these supervisory roles depends on whether human response and operator workload fit within the intervention window. NASA's detect-and-avoid (DAA) simulations report about 17 to 22 s to begin responding and 34 to 42 s to upload a resolution \citep{fern2015daa}. The routine multi-aircraft operations working group uses intervention windows, switching times, and service times to examine supervisory arrangements \citep{emerson2026ramao}. Queuing models address the additional delay caused by shared radio or operator resources \citep{saephan2025queuing,cummings2014task}. Low utilization can also impair vigilance \citep{cummings2013boredom}. These findings require the model to distinguish the latency of the communication link from the time the human needs to execute the intervention.

Crediting such a response in a safety assessment requires evidence that the human can act under the assumed conditions. The probability of one that the FAA and JARUS guidance allow for a reasonable crew task is a conditional assessment convention \citep{faa2011ac231309,jarus2015amc1309}. It does not establish that an operator reached through an imperfect link will receive and complete an action in time. The National Research Council's warning against assigning humans an unrestricted rescue role motivates explicit analysis of those conditions \citep{nrc2014autonomy}.

Those intervention assumptions also distinguish architectures that retain human backup from architectures designed to operate without it. Reduced crew concepts retain ground support and couple its requirements with cockpit automation and data links \citep{matessa2017rco,bailey2017assessment}. Digital Flight instead considers automated self separation without human backup \citep{wing2022digital,lacher2025viability1}. Its Required Conflict Management Performance concept links integrated subsystem performance to separation requirements and a safety outcome \citep{wing2025viability2}. That report develops the performance concept qualitatively and leaves the quantitative allocation outside its scope. The present model addresses functions that retain remote human intervention and evaluates the safety benefit that this intervention can supply.

\subsection{Communication Performance and Joint Safety Assessment}
For architectures that retain remote intervention, the requirements on the communication subsystem must support the human role just described. The Required Communication Performance and Required C2 Performance concepts named in the Introduction specify transaction time, continuity, availability, and integrity, and the second acknowledges the dependence on onboard automation \citep{icao2017pbcs,jarus2016rlp}. Spectrum studies and minimum C2 system standards support link design within that framework \citep{itu2011m2233,rtca2023do377b}. The Specific Operations Risk Assessment addresses link characteristics and recovery from human error through separate safety objectives \citep{jarus2024sora}. These documents provide requirements and assessment categories without expressing the level-of-safety as a function of the frequency, urgency, and complexity of interventions.

Assessing whether a link can meet these requirements draws on channel measurements and operational experiments. NASA's C2 work addresses line of sight and satellite links, including propagation conditions and airframe shadowing \citep{kerczewski2017progress,matolak2017air,sun2017air}. Controller simulations measure operational effects of delayed exchanges, while lost link assessments examine contingency procedures \citep{hayashi2022effects,comstock2015uas,sakakeeny2024initial}. Cellular requirements also vary tolerable latency with the control mode \citep{abdalla2021communications}. The FAA's 2025 proposal for operations beyond visual line of sight calls for assessment of latency and intermittent lost link \citep{faa2025bvlos}. Such evidence helps specify the link and its operating environment. A model of the level-of-safety additionally requires the frequency, urgency, and complexity of the interventions that use it.

Connecting these link constraints to autonomy has motivated studies of joint communication and system performance. The simulation studies described in the Introduction measure safety or mission outcomes as the communication link deteriorates, with the control architecture held fixed \citep{pang2026reliability,bulusu2022impact}. \citet{safwat2025delays} combine human response, trajectory negotiation, and communication delay in a common timing budget. \citet{tabassum2022dynamic} determine when onboard detect and avoid must act without waiting for delayed remote control. These analytical studies connect autonomy and latency without the MTBI or the availability of the link.

\subsection{Risk Models for Unsuccessful Intervention}
The joint model missing from these studies can build on established human reliability models of response against a deadline. Time reliability correlations relate unsuccessful action to the time available for a response \citep{hall1982post}. The NASA PRA guide treats both the time available and the time required as random quantities \citep{nasa2011pra}. Its flight termination example requires a remote officer to act before a failing launch vehicle becomes hazardous. This supplies the timing structure used here. The availability and the latency of the link add the communication conditions under which that response can succeed.

Converting this timing probability into a failure rate also requires a model of intervention frequency and unsuccessful intervention. Reliability analysis of diverse channels examines how one channel can mitigate failures of another under stated dependence assumptions \citep{littlewood2012reasoning}. \citet{alves2018considerations} apply related reasoning to monitored autonomous aircraft and stringent safety objectives. The difficulty of demonstrating very low software failure rates through testing motivates such architectural reasoning \citep{butler1993infeasibility,littlewood1993validation}. Automated driving studies reach a related conclusion about the exposure needed for statistical demonstration \citep{kalra2016driving}. Remote intervention can relax the required MTBI only to the extent that its own failure probability is bounded.

The resulting failure rate also depends on how uncertainty about the autonomy parameters is represented. The probability of frequency approach represents uncertainty about an event rate through a distribution over that rate \citep{kaplan1981risk,nasa2011pra}. Under this interpretation, a mean risk and a high quantile answer different assurance questions. Applying the approach to uncertain autonomy parameters allows compliance with a safety budget to be stated at a chosen assurance level. The resulting statement remains conditional on the distributions and dependence assumptions used to represent those parameters.

The frequency and timing mechanisms underlying this calculation also appear in the other partially autonomous machine systems named in the Introduction, which are road vehicles, ships, and crewed spacecraft \citep{hickert2023cooperation,rodseth2022towards,frank2013autonomous,valinia2022safe}. Debris avoidance also involves a ground decision and command chain before a possible close approach \citep{bastidavirgili2019cream,nasa2026cahandbook}. These applications share the dependence on timely external assistance. Their exposure units, hazard definitions, and safety budgets require application specific choices.

\subsection{Research Gap}
Taken together, this literature covers various components of a safety calculation but leaves the joint model unresolved. In particular, the cited coupling studies do not jointly include the MTBI, the urgency, the complexity, the availability, and the latency in a level-of-safety concept. We formulate that model for interventions that arise independently and do not queue for the operator. It then supports constant risk comparisons and uncertainty analysis for specified functions, while leaving broader human factors and certification decisions outside the model.

\section{Methodology \label{sec: methodologies}}
To construct the model, we first define the autonomy and communication parameters in Section~\ref{subsec: parameterization}. Sections~\ref{subsec: success probability} and~\ref{subsec: residual risk} derive the intervention success probability and the residual failure rate that expresses the level-of-safety. Section~\ref{subsec: monte carlo} propagates the uncertainty in the autonomy parameters, and Section~\ref{subsec: multi phase} extends the calculation across flight phases. Section~\ref{subsec: pareto} then introduces the criticality ratio and derives the constant risk frontiers.

\subsection{Performance of the Autonomy and Communication Subsystems\label{subsec: parameterization}}
Inspired by the paradigm shift in the way navigational subsystems are prescribed, i.e., when in the late 1980's the aviation community went from prescribing the components and parameters of the navigation subsystem to prescribing a level of performance that had to be demonstrated via analysis or testing, we prescribe the autonomy and communication subsystems in terms of required performance parameters.
The autonomy parameters describe how often an intervention is required, how long the intervention window is, and how long it takes to complete the intervention. The communication parameters describe whether the link is available when the intervention is required and the latency it adds. Together, these quantities determine whether a remote intervention succeeds.

Within this setting, three parameters characterize the performance of the autonomy subsystem,
\begin{equation}
    \theta_A = \bigl(\mathrm{MTBI}, \mathcal{U}, \mathcal{C}\bigr),
    \label{eq:theta_A}
\end{equation}
where the components have the following meanings,
\begin{itemize}
\item \emph{Mean time between interventions.} The $\mathrm{MTBI}$ [time] is the mean interval between required interventions. We assume exponentially distributed intervals. The intervention frequency is $\mathcal{F}=1/\mathrm{MTBI}$, and we call each instance in which an intervention is required a demand. We define $\alpha$ through $\mathrm{MTBI}=10^{\alpha}$ operating hours, so one unit of $\alpha$ corresponds to one order of magnitude in the MTBI. Aviation uses flight hours as the exposure unit. The scale describes a specified function, alerting rule, and operating setting. It measures how often intervention is required and does not specify which decisions the human or machine controls. It is also only one of the three autonomy parameters. With the same MTBI, an autonomy subsystem with an average urgency of 1 s is not equivalent to an autonomy subsystem with an average urgency of 10 s.
\item \emph{Urgency.} The urgency $\mathcal{U}$ [time] is the mean time from the moment an intervention is deemed necessary to the moment failure is certain if no intervention is made. We assume this time is exponentially distributed. Lower $\mathcal{U}$ denotes a more urgent need for intervention. The function, operating conditions, and alert timing determine the urgency. We assume that there is no trend in $\mathcal{U}$ with $\alpha$.
\item \emph{Complexity.} The complexity $\mathcal{C}$ [time] is the mean time needed, after the request for help reaches the operator, to understand the situation, then decide on and execute the intervention. We assume the total time to complete these activities is exponentially distributed. The numerical illustration assumes that this mean decreases with $\alpha$. That trend requires measurement for a particular design and function.
\end{itemize}

Complementing this autonomy triple, two parameters characterize the performance of the communication subsystem,
\begin{equation}
    \theta_C = (P_A,\ \varepsilon),
    \label{eq:theta_C}
\end{equation}
where the parameters are defined as follows.
\begin{itemize}
\item \emph{Communication availability.} The availability $P_A$ is the probability that the communication link is available when an intervention is required. Continuity describes whether an available link remains usable during the exchange \citep{pang2026reliability}. The base model represents availability at demand and does not model continuity separately.
\item \emph{Communication latency.} The latency $\varepsilon$ [time] is the total time delay that the link introduces in the request downlink and the corrective command uplink.
\end{itemize}

To keep these parameter groups separate, we characterize $\theta_A$ with an idealized link and then vary $\theta_C$ independently. Their joint effect enters through the intervention success probability in Section~\ref{subsec: success probability}.

The autonomy parameters characterized in this way are means of random intervals, and those intervals already describe the variation between encounters. The means themselves remain uncertain because the analyst has incomplete knowledge of a design's capability. Following the probability of frequency approach \citep{kaplan1981risk,nasa2011pra}, we represent this epistemic uncertainty by a distribution over the autonomy parameters,
\begin{equation}
    \theta_A = \bigl(\mathrm{MTBI}, \mathcal{U}, \mathcal{C}\bigr)
    \sim
    \pi_A(\theta_A \mid \alpha),
    \label{eq:capability_distribution}
\end{equation}
so each draw supplies one MTBI, one urgency, and one complexity.

We represent this uncertainty through the frequency $\mathcal{F}$ because the failure rate depends on the MTBI through its reciprocal. A frequency draw fixes the corresponding MTBI. Under this distribution, $10^{\alpha}$ is the reciprocal of the mean frequency, which we call the nominal MTBI. It is not the mean of the sampled MTBIs, which is infinite under an exponential frequency distribution. We write the means as $m_{\mathcal{F}}(\alpha)$, $m_{\mathcal{U}}$, and $m_{\mathcal{C}}(\alpha)$. Thus, $m_{\mathcal{F}}(\alpha)=10^{-\alpha}$ per flight hour.

To specify this capability distribution for illustration, we use exponential marginals for $\mathcal{F}$, $\mathcal{U}$, and $\mathcal{C}$. This is the maximum entropy family on nonnegative values with a specified mean \citep{jaynes1957information}. It also fixes the coefficient of variation at one. The illustration therefore represents diffuse parameter uncertainty, not an uncertainty distribution fitted to operational evidence. We assume a decreasing mean complexity as $\alpha$ increases and hold mean urgency fixed for the function under study.

The diffuse uncertainty represented by these marginals can be narrowed as measured evidence accumulates. For example, a gamma distribution for the frequency can represent increasing information from observed interventions. A joint distribution can also encode dependence between frequency, urgency, and complexity. A point mass recovers the deterministic parameter model. The derivations below condition on $\theta_A$, while sampling uses lowercase $f$, $u$, and $c$ for its realized components. Table~\ref{tab:parameters} summarizes the notation.

\begin{table}[htbp]
\centering
\caption{Notation of the model. The parameters are grouped into the autonomy triple $\theta_A$ and the communication pair $\theta_C$, followed by the derived quantities and the requirements. Lowercase $f$, $u$, and $c$ are samples of $\mathcal{F}$, $\mathcal{U}$, and $\mathcal{C}$.}
\label{tab:parameters}
\footnotesize
\begin{tabular}{l p{0.42\textwidth} l l}
\hline
Symbol & Definition & Dimension & Category \\
\hline
$\mathrm{MTBI}$ & Mean time between interventions (nominal value $10^{\alpha}$ flight hours) & time & Autonomy \\
$\mathcal{F}$ & Intervention frequency, $\mathcal{F}=1/\mathrm{MTBI}$ & time$^{-1}$ & Autonomy (derived) \\
$\alpha$ & Logarithm of the nominal MTBI in flight hours & none & Autonomy (derived) \\
$\mathcal{U}$ & Intervention urgency (mean time to failure without intervention), which depends on the function or flight phase & time & Autonomy \\
$\mathcal{C}$ & Intervention complexity (mean time to complete intervention with no latency) & time & Autonomy \\
$P_A$ & Communication availability & none & Communication \\
$\varepsilon$ & Communication latency & time & Communication \\
$\pi_A(\theta_A \mid \alpha)$ & Capability distribution of the autonomy triple & none & Autonomy \\
$m_{\mathcal{F}},\,m_{\mathcal{U}},\,m_{\mathcal{C}}$ & Means of $\mathcal{F}$, $\mathcal{U}$, and $\mathcal{C}$ under the capability distribution & as above & Autonomy \\
$\rho$ & Criticality ratio, $\rho=(\mathcal{C}+\varepsilon)/\mathcal{U}$ & none & Derived \\
$P_{\mathrm{int}}$ & Probability that an intervention succeeds & none & Derived \\
$R$ & Residual failure rate for given $\theta_A$ and $\theta_C$, which expresses the level-of-safety & time$^{-1}$ & Derived \\
$\bar{R}$ & Mean of $R$ under the capability distribution & time$^{-1}$ & Derived \\
$R_{\mathrm{budget}}$ & Safety budget (allowable failure rate) & time$^{-1}$ & Requirement \\
$\delta$ & Tolerated probability of exceeding the budget & none & Requirement \\
\hline
\end{tabular}
\end{table}

With a frequency drawn from this capability distribution, exponential intervals give a conditional Poisson intervention process. The number of demands over $t$ flight hours therefore satisfies
\begin{equation}
    \mathbb{P}\bigl(N(t)=k \mid \mathcal{F}\bigr)
    =
    \frac{(\mathcal{F}t)^{k}\,e^{-\mathcal{F}t}}{k!},
    \qquad k = 0, 1, 2, \dots,
    \label{eq:poisson_count}
\end{equation}
with $t=1$ giving the count per flight hour.

Averaging this conditional process over the exponential frequency distribution gives the unconditional count distribution,
\begin{equation}
    \mathbb{P}\bigl(N(t)=k\bigr)
    =
    \frac{1}{1+10^{-\alpha}t}
    \left(\frac{10^{-\alpha}t}{1+10^{-\alpha}t}\right)^{k},
    \qquad k = 0, 1, 2, \dots,
    \label{eq:geometric_count}
\end{equation}
which has mean $10^{-\alpha}t$ and a heavier tail than a Poisson count with that mean. The additional spread comes from uncertainty about the frequency. The conditional Poisson process remains the basis for the failure rate derivation.

\subsection{Intervention Success Probability\label{subsec: success probability}}
Each time an intervention is deemed necessary, the intervention will only succeed if it is completed before the intervention window runs out. Set $t=0$ at the time the intervention is requested, conditional on $\theta_A$. Let $W$ be the intervention window, i.e., the time to failure without completed intervention; and $D$ the nominal completion time, i.e., the time to complete the intervention if there is no latency. We take $W$ and $D$ as exponential with means $\mathcal{U}$ and $\mathcal{C}$, respectively. The communication link adds the latency $\varepsilon$. The intervention arrives in time when,
\begin{equation}
    \varepsilon + D \;\leq\; W.
    \label{eq:success_condition}
\end{equation}

To calculate this success probability, we represent the nominal completion time by its cumulative distribution,
\begin{equation}
    F_{D}\bigl(t;\,\mathcal{C}\bigr)
    =
    \mathbb{P}\bigl(D \leq t\bigr)
    =
    1-e^{-t/\mathcal{C}},
    \qquad t \geq 0,
    \label{eq:delay_cdf}
\end{equation}
which gives the probability of completing the intervention within a specified time. The variable $t$ here measures the time to initiate and complete an intervention assuming no latency and is separate from mission exposure. Comparing this distribution with that of $W$ accounts for different urgencies. Other distributions for the nominal completion time may be used without changing the success condition.

Combining this distribution with an independent $W$ and the availability of the link gives the success probability,
\begin{align}
    P_{\mathrm{int}}(\theta_A, \theta_C)
    &=
    P_A\,
    \mathbb{P}\bigl(\varepsilon + D \leq W\bigr)
    \nonumber \\
    &=
    P_A
    \int_0^{\infty}
    \frac{e^{-t/\mathcal{C}}}{\mathcal{C}}\,
    e^{-(t+\varepsilon)/\mathcal{U}}\,\mathrm{d}t
    \nonumber \\
    &=
    P_A\,
    \frac{e^{-\varepsilon/\mathcal{U}}}{1 + \mathcal{C}/\mathcal{U}},
    \label{eq:p_int}
\end{align}
where an outage present at the demand is assumed to outlast the intervention window. This assumption excludes an intervention after the link becomes available again and is conservative for short outages. Section~\ref{subsec: pareto} examines that extension. Interruptions during an exchange are not modeled separately.

Within this success probability, the factor $e^{-\varepsilon/\mathcal{U}}$ represents survival during the latency. The factor $1/(1+\mathcal{C}/\mathcal{U})$ is the probability that the intervention is completed before failure. Their product follows from the memoryless exponential intervention window. Independence between $W$ and $D$ separates the external failure dynamics from the completion of the intervention. Applications where harder situations both shorten the intervention window and lengthen the nominal completion time require a joint timing model.

Changing the nominal completion time distribution replaces the completion factor with its Laplace transform, $\mathcal{L}_D(1/\mathcal{U})$. Changing the intervention window distribution also changes the latency effect. The exponential intervention window allows arbitrarily short values. It can therefore overstate the penalty of a small latency when an alerting function guarantees minimum lead time.

With the exponential timing assumptions retained, exact success depends on two dimensionless groups, $\mathcal{C}/\mathcal{U}$ and $\varepsilon/\mathcal{U}$. Allocations with the same two ratios have the same normalized success probability. Success increases with the availability and with a longer $\mathcal{U}$, and decreases with the complexity and the latency. As $\mathcal{U}\to\infty$, success tends to $P_A$ under the persistent outage assumption. With finite outages and an indefinitely long intervention window, eventual link restoration would instead permit success probability to approach one.

These timing probabilities assume that an operator can begin responding without waiting for another intervention to finish. With consistent time units, utilization is $\eta=\mathcal{F}\mathcal{C}$. The non-overlapping approximation requires $\eta\ll1$. When several aircraft share an operator, the aggregate intervention frequency must enter this check.

If interventions must queue with Poisson arrivals and exponential service, their mean nominal completion time becomes $\mathcal{C}/(1-\eta)$ \citep{cummings2014task}. Stability requires $\eta<1$, but that condition alone does not justify neglecting queuing. We retain the uncongested approximation. Shared supervision and the vigilance effects of very low utilization require additional human performance modeling \citep{cummings2013boredom}.

\subsection{Level-of-Safety and Safety Budget\label{subsec: residual risk}}
Under the uncongested assumption, the intervention frequency and the probability of an unsuccessful intervention determine the residual failure rate. Demands occur at rate $\mathcal{F}=1/\mathrm{MTBI}$. Each intervention fails independently with probability $1-P_{\mathrm{int}}$. Because $W$ is the time to the selected system failure event without a successful intervention, an unsuccessful intervention produces that event. The conditional residual failure rate is
\begin{equation}
    R(\theta_A, \theta_C)
    =
    \mathcal{F}
    \left[1-P_{\mathrm{int}}(\theta_A, \theta_C)\right]
    =
    \mathcal{F}
    \left[
    1-P_A\,\frac{e^{-\varepsilon/\mathcal{U}}}{1 + \mathcal{C}/\mathcal{U}}
    \right].
    \label{eq:residual_risk}
\end{equation}
Interpreting this residual failure rate requires a safety budget for the same failure event and exposure unit. Applications with several severity classes require separate event definitions or an explicit model relating unsuccessful intervention to each outcome.

At the process level, independent thinning gives a Poisson failure process with this rate $R$. Over $t$ flight hours, its mean count is $Rt$ and its probability of no failure is $e^{-Rt}$. We use this residual failure rate as the level-of-safety of the selected function. For fixed parameters, it meets a target level-of-safety expressed in failures per flight hour when
\begin{equation}
    R(\theta_A, \theta_C) \;\leq\; R_{\mathrm{budget}},
    \label{eq:budget_fixed}
\end{equation}
where $R_{\mathrm{budget}}$ is the safety budget allocated to the selected failure event.

The limiting cases of Eq.~\eqref{eq:residual_risk} identify the subsystem that limits the level-of-safety.
\begin{itemize}
    \item \emph{Perfect communication.} If $P_A \to 1$ and $\varepsilon \to 0$, the failure rate tends to $\mathcal{F}\,\frac{\mathcal{C}/\mathcal{U}}{1+\mathcal{C}/\mathcal{U}}$ and is set entirely by the autonomy triple. A complex or urgent intervention (large $\mathcal{C}/\mathcal{U}$) leaves a residual failure rate even over a perfect link.
    \item \emph{Instantaneous intervention.} With $\mathcal{C} \to 0$ and negligible latency, $\varepsilon/\mathcal{U} \to 0$, the rate tends to $\mathcal{F}\,(1-P_A)$, and the residual failure rate is set entirely by the availability.
    \item \emph{Latency dominated link.} With $\varepsilon \gg \mathcal{U}$, a latency long relative to the urgency, $e^{-\varepsilon/\mathcal{U}} \to 0$ and $R \to \mathcal{F}$ regardless of $P_A$ and $\mathcal{C}$. The remote intervention rarely arrives in time, so the safety budget must be met by the autonomy subsystem alone, that is, by reducing the intervention frequency.
    \item \emph{Full autonomy.} With $\mathcal{F} \to 0$, the rate $R \to 0$ for any communication performance. The link ceases to be safety critical, although it may still be needed for coordination or efficiency.
\end{itemize}
\subsection{Monte Carlo Risk Evaluation\label{subsec: monte carlo}}
For this conditional failure rate, only uncertainty about the autonomy parameters remains to be evaluated numerically. Replication $i$ draws a triple $\bigl(f^{(i)},u^{(i)},c^{(i)}\bigr)$ from the capability distribution. The default marginals are
\begin{equation}
\begin{aligned}
    f^{(i)} &\sim
    \mathrm{Exp}\bigl(m_{\mathcal{F}}(\alpha)\bigr),\quad
    u^{(i)} \sim
    \mathrm{Exp}\bigl(m_{\mathcal{U}}\bigr), \quad
    c^{(i)} \sim
    \mathrm{Exp}\bigl(m_{\mathcal{C}}(\alpha)\bigr),
\end{aligned}
\label{eq:component_sampling}
\end{equation}
where $\mathrm{Exp}(m)$ denotes an exponential distribution with mean $m$.

These draws use $m_{\mathcal{F}}(\alpha)=10^{-\alpha}$ per flight hour, a decreasing $m_{\mathcal{C}}(\alpha)$, and a fixed $m_{\mathcal{U}}$ for the selected function. The default draws are independent. A joint distribution is needed when parameter dependence is supported by evidence. Substitution in Eq.~\eqref{eq:residual_risk} gives a conditional rate for each draw, and their average estimates
\begin{align}
    \bar{R}(\alpha;\theta_C)
    &=
    \mathbb{E}_{\theta_A \sim \pi_A(\cdot \mid \alpha)}
    \bigl[R(\theta_A, \theta_C)\bigr]
    \approx
    \frac{1}{N_s}
    \sum_{i=1}^{N_s}
    R\bigl(\theta_A^{(i)}, \theta_C\bigr)
    \nonumber \\
    &=
    \frac{1}{N_s}
    \sum_{i=1}^{N_s}
    f^{(i)}
    \left[
    1-P_A\,
    \frac{e^{-\varepsilon/u^{(i)}}}
    {1+c^{(i)}/u^{(i)}}
    \right],
    \label{eq:mc_estimator}
\end{align}
in which the expectation is over uncertainty in the autonomy parameters.

The sampled mean accounts for nonlinearity that a rate evaluated at mean parameters would miss. The frequency scales the rate, while the urgency and the complexity enter through the success probability. With independent components, $\bar{R}=m_{\mathcal{F}}(\alpha)[1-P_A\mathbb{E}(q)]$, where $q=P_{\mathrm{int}}/P_A$. Nonlinearity in urgency and complexity means that the rate at the mean parameters generally differs from the mean rate. The direction of that difference is not fixed for all parameter distributions.

At zero latency, this sampled mean has an analytical check in terms of $a=m_{\mathcal{C}}/m_{\mathcal{U}}$. The independent exponential capability marginals give $\mathbb{E}(q)=(1-a+a\ln a)/(1-a)^2$. The continuous limit at $a=1$ is $1/2$. This expression provides a check on the sampler. In the numerical illustrations, short sampled urgencies raise the mean residual failure rate above the rate at the mean parameters.

Beyond the mean, these sampled rates support an assurance statement over uncertain design capability. We define assurance as the probability, over the specified parameter uncertainty, that the allocation meets its safety budget. The requirement is
\begin{equation}
    \mathbb{P}_{\theta_A \sim \pi_A(\cdot \mid \alpha)}
    \bigl[
    R(\theta_A, \theta_C) \leq R_{\mathrm{budget}}
    \bigr]
    \;\geq\;
    1-\delta,
    \label{eq:safety_budget}
\end{equation}
where $\delta$ is the tolerated probability of exceeding the budget under the capability distribution.

Evaluating this assurance requirement uses the empirical $(1-\delta)$ quantile, with rank $\lceil(1-\delta)N_s\rceil$ in the ordered rates. Compliance requires this quantile to remain within the budget. Expected value compliance instead requires $\bar{R}\le R_{\mathrm{budget}}$. These are different criteria for the same allocation. Common random numbers across candidate designs reduce sampling noise in their comparison.

\subsection{Risk Across Multiple Flight Phases\label{subsec: multi phase}}
The same risk and assurance calculations can be extended to missions whose autonomy and communication parameters vary over time. Across a mission, the intervention frequency and the urgency can vary with flight phase. Availability and latency can also vary with coverage and route. Letting the capability distribution and communication parameters vary with time gives
\begin{equation}
    N_H
    =
    \int_0^{T_m}
    \bar{R}\bigl(\alpha(t),t;\theta_C(t)\bigr)\,\mathrm{d}t
    \;\leq\;
    N_{\mathrm{budget}},
    \label{eq:mission_risk}
\end{equation}
where $T_m$ is mission duration, $N_H$ the expected number of modeled failures, and $N_{\mathrm{budget}}$ the corresponding mission budget. The explicit argument $t$ marks the dependence of the capability distribution on the flight phase.

With piecewise constant parameters, this mission integral becomes a sum over flight phases or route segments. Each segment retains the conditional failure rate of Eq.~\eqref{eq:residual_risk}. Mission assurance requires joint sampling because uncertainty about the same design can persist across phases. The remaining illustration uses stationary parameters.

\subsection{Criticality Ratio and the Constant Risk Frontier\label{subsec: pareto}}
Within a specified operating setting, the exact success probability depends on $\varepsilon/\mathcal{U}$ and $\mathcal{C}/\mathcal{U}$. Their sum is the mean time required to understand, decide on, communicate and execute an intervention relative to the urgency,
\begin{equation}
    \rho = \frac{\mathcal{C}+\varepsilon}{\mathcal{U}},
    \label{eq:rho}
\end{equation}
which we call the criticality ratio. A small ratio means that an intervention uses little of the time that the urgency allows.

\begin{figure}[htbp]
\centering
\includegraphics{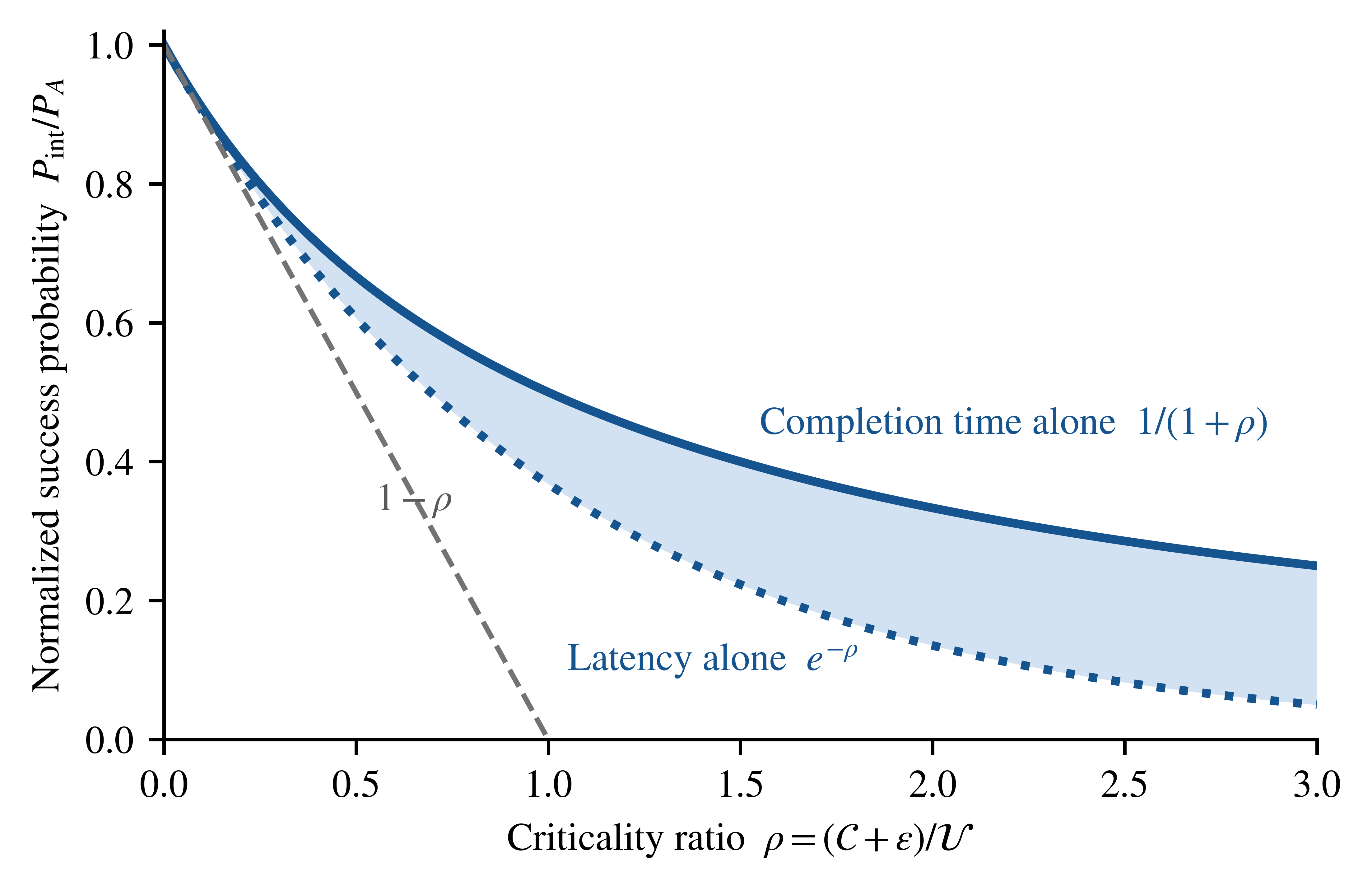}
\caption{Normalized intervention success probability against $\rho=(\mathcal{C}+\varepsilon)/\mathcal{U}$. The band covers every division of the total time between latency and an exponential nominal completion time. Its bounds are $e^{-\rho}$ and $1/(1+\rho)$, both equal to $1-\rho$ to first order for small $\rho$.}
\label{fig: rho_success}
\end{figure}

The ratio determines the normalized success probability, and through it the shape of the tradeoff between the MTBI and the availability, up to the share of the latency in the total time, $s=\varepsilon/(\mathcal{C}+\varepsilon)$. The exact normalized success probability is $q=e^{-s\rho}/[1+(1-s)\rho]$. It decreases with the latency share, so for every division of the total time it satisfies
\begin{equation}
    e^{-\rho}
    \;\le\;
    \frac{P_{\mathrm{int}}}{P_A}
    \;\le\;
    \frac{1}{1+\rho}.
    \label{eq:rho_bounds}
\end{equation}
These bounds correspond to latency alone and to an exponential nominal completion time alone, respectively. For the same total time, an exponential nominal completion time allows some interventions to finish early. Both bounds equal $1-\rho$ to first order as $\rho\to0$, so the band is narrow where remote intervention is most effective. Figure~\ref{fig: rho_success} shows the resulting band.

The upper bound is the value of $q$ at zero latency share, which we write $q_0(\rho)=1/(1+\rho)$. It expresses the tradeoff through the ratio alone and is used for the frontier charts. This treats the total time as exponential with mean $\mathcal{C}+\varepsilon$. It is exact at zero latency and optimistic for a positive latency share. At $\rho=0.1$, using the lower bound raises the required MTBI for unit availability by about 5\%. At $\rho=0.9$, the increase is about a quarter. The exact Eq.~\eqref{eq:p_int} remains the basis of the numerical risk maps.

With this bound and fixed values of the urgency, the complexity, and the latency, solving Eq.~\eqref{eq:budget_fixed} with equality for the required MTBI gives the constant risk frontier,
\begin{equation}
    \mathrm{MTBI}
    =
    \frac{1-P_A\,q_0(\rho)}{R_{\mathrm{budget}}}.
    \label{eq:iso_frontier}
\end{equation}

This frontier is the Pareto tradeoff between the MTBI and the availability at a constant level-of-safety. Along it, a lower MTBI can be accepted only with a higher availability. The safety budget sets the scale of the frontier, and the ratio prescribes its shape through $q_0(\rho)$.

\begin{figure}[htbp]
\centering
\includegraphics{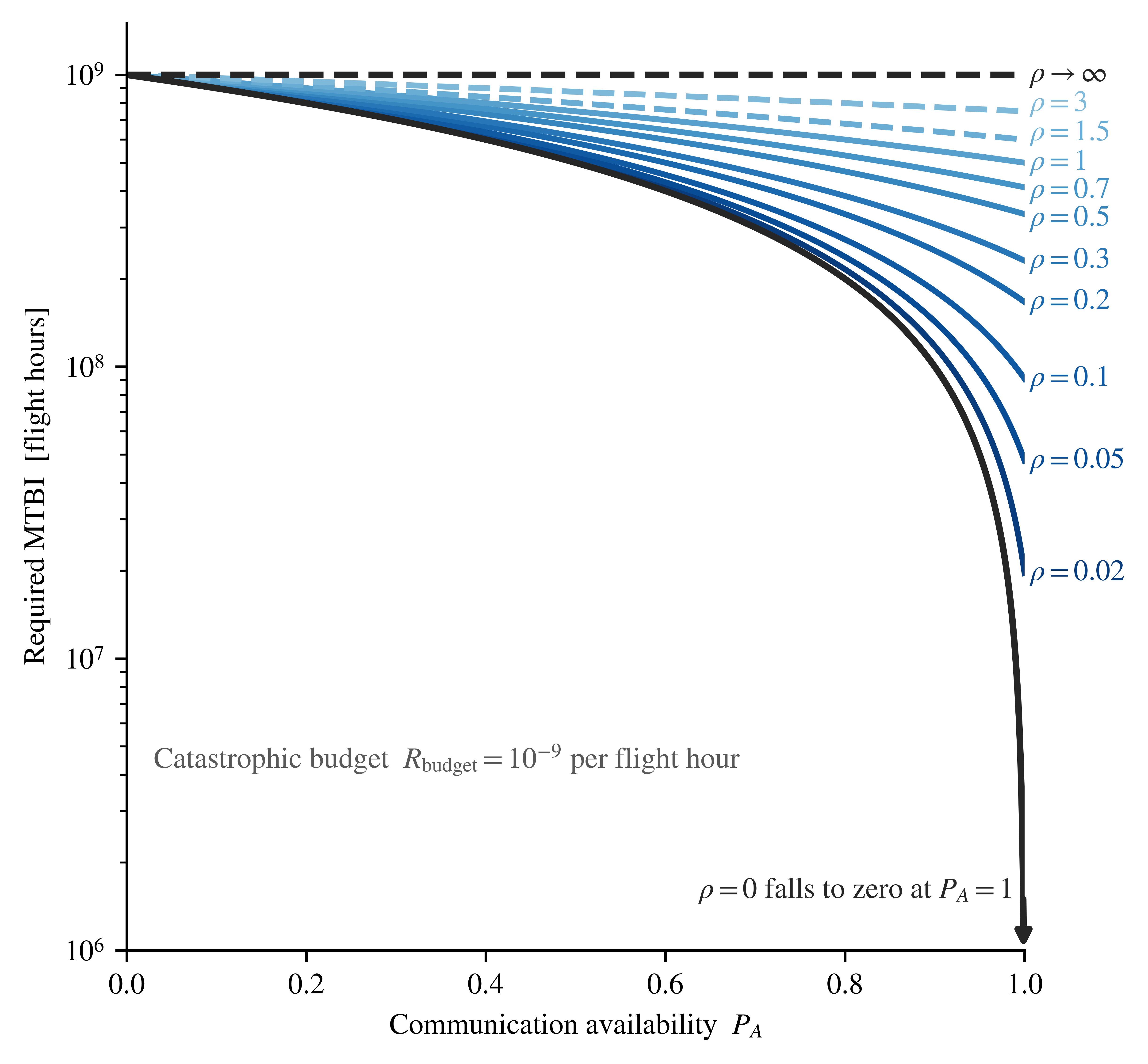}
\caption{Constant risk frontiers at $R_{\mathrm{budget}}=10^{-9}$ per flight hour using $q_0(\rho)=1/(1+\rho)$. Allocations above a curve meet the budget under this bound, which is exact at zero latency and optimistic otherwise. Small ratios permit greater MTBI relief from improved availability. Dashed curves have $\rho>1$. The frontier for $\rho=0$ falls to zero at unit availability and leaves the plotted range. The curves assume that an outage outlasts the intervention window and rescale inversely with the budget.}
\label{fig: pareto_rho}
\end{figure}

The resulting frontiers in Figure~\ref{fig: pareto_rho} identify MTBI and availability combinations meeting a safety budget of $10^{-9}$ per flight hour. Allocations above a curve meet the budget. At $P_A=0$, the required MTBI is $1/R_{\mathrm{budget}}$, since remote intervention is unavailable. All curves rescale inversely with the budget. For example, a budget of $10^{-7}$ per flight hour lowers them by a factor of one hundred.

At unit availability, these frontiers reduce the required MTBI by $(1+\rho)/\rho$ relative to operation without remote intervention. The factors are about one hundred at $\rho=0.01$, about fifty at $\rho=0.02$, eleven at $\rho=0.1$, and two at $\rho=1$. These values are exact for a perfect link and fixed autonomy parameters. For a positive latency they bound the relief from above at the same $\rho$. As $\rho\to\infty$, remote intervention becomes ineffective and the frontier tends to $\mathrm{MTBI}=1/R_{\mathrm{budget}}$. At the opposite limit of $\rho=0$, the frontier is $\mathrm{MTBI}=(1-P_A)/R_{\mathrm{budget}}$. It falls to zero at unit availability, so it has no endpoint on the logarithmic axis and is drawn with an arrow.

The slope of these frontiers quantifies the value of additional availability. Under the upper bound, it is $\mathrm{d}\ln\mathrm{MTBI}/\mathrm{d}P_A=-q_0/(1-P_Aq_0)$. Its magnitude increases from $q_0$ at zero availability to $1/\rho$ at unit availability. The final increments of availability therefore have the greatest effect when $\rho$ is small. Halving the required MTBI needs $P_A=(1+\rho)/2$. At $\rho=1$, only unit availability achieves that reduction. Above one, even the upper success bound cannot halve the requirement.

The relief quantified above depends on the base assumption that an outage persists throughout the intervention window. Consider instead exponential up and down periods with means $T_{\mathrm{up}}$ and $\tau$. Their stationary availability is $P_A=T_{\mathrm{up}}/(T_{\mathrm{up}}+\tau)$. A demand arriving during an outage can wait for restoration. The probability of surviving this additional wait is $1/(1+\tau/\mathcal{U})$. Writing $\kappa=T_{\mathrm{up}}/\mathcal{U}$ gives an effective availability,
\begin{equation}
    P_{\mathrm{eff}}
    =
    P_A + \frac{P_A\,(1-P_A)}{P_A+\kappa\,(1-P_A)},
    \label{eq:p_eff}
\end{equation}
which replaces $P_A$ in Eq.~\eqref{eq:iso_frontier} and tends to $P_A$ as $\kappa\to\infty$.

Allowing this restoration increases intermediate effective availabilities without changing the endpoints at zero and one. The persistent outage curves therefore give larger MTBI requirements than this restoration model. At $\rho=0.1$, the largest reduction in the required MTBI over all availabilities is about 6\% for $\kappa=100$ and 54\% for $\kappa=1$. These comparisons concern waiting for restoration before an exchange. They do not resolve interruptions after an exchange has begun.

Together, these timing and outage results delimit how much a better communication subsystem can compensate for a lower MTBI. For small $\rho$ and small unavailability, the approximate frontier is $\mathrm{MTBI}\approx[(1-P_A)+\rho]/R_{\mathrm{budget}}$. Unavailability and late completion then consume the same failure probability allowance. As $\rho$ approaches one, even the upper bound gives limited relief. The exact tradeoff also depends on the latency share within the bounds of Eq.~\eqref{eq:rho_bounds}. These frontiers hold the urgency, the complexity, and the latency fixed while varying the MTBI and the availability. If the complexity changes with $\alpha$, or the autonomy parameters are uncertain, the full model must be evaluated along that design relation.

\section{Numerical Illustration\label{sec: results}}
The numerical illustration applies the model to two functions of a remotely piloted aircraft with assumed parameter values. At each $\alpha$, we draw $N_s=40{,}000$ triples from the exponential capability marginals. Each draw gives a conditional rate through Eqs.~\eqref{eq:p_int} and~\eqref{eq:residual_risk}. We summarize the rates by their mean and 95th percentile, corresponding to $\delta=0.05$. Explicit sampling of $W$ and $D$ reproduces the closed form within sampling error.

We compare these sampled rates with the aviation safety budgets of $10^{-3}$, $10^{-5}$, $10^{-7}$, and $10^{-9}$ per flight hour, drawn from FAA and EASA severity guidance \citep{faa2011ac231309,easa2023cs25}. We treat each as a safety budget for the modeled failure event. The frequency mean is $m_{\mathcal{F}}(\alpha)=10^{-\alpha}$ per flight hour. For the first function, we assume that the logarithm of mean complexity decreases linearly with $\alpha$, from 30~s at $\alpha=-2$ to 2~s at $\alpha=10$. The second function has a fixed mean complexity of its own. Mean urgency is specified separately for each function. Section~\ref{sec: evaluation} describes how measurements would replace these assumptions.

With these assumptions, the first chart varies $\alpha$ and the availability while holding the urgency and the latency fixed. The second fixes a low intervention frequency and varies the performance of the communication subsystem for a function with a short urgency. These choices separate the effect of requiring fewer interventions from the effect of delivering each intervention sooner.

\begin{figure}[htbp]
\centering
\includegraphics{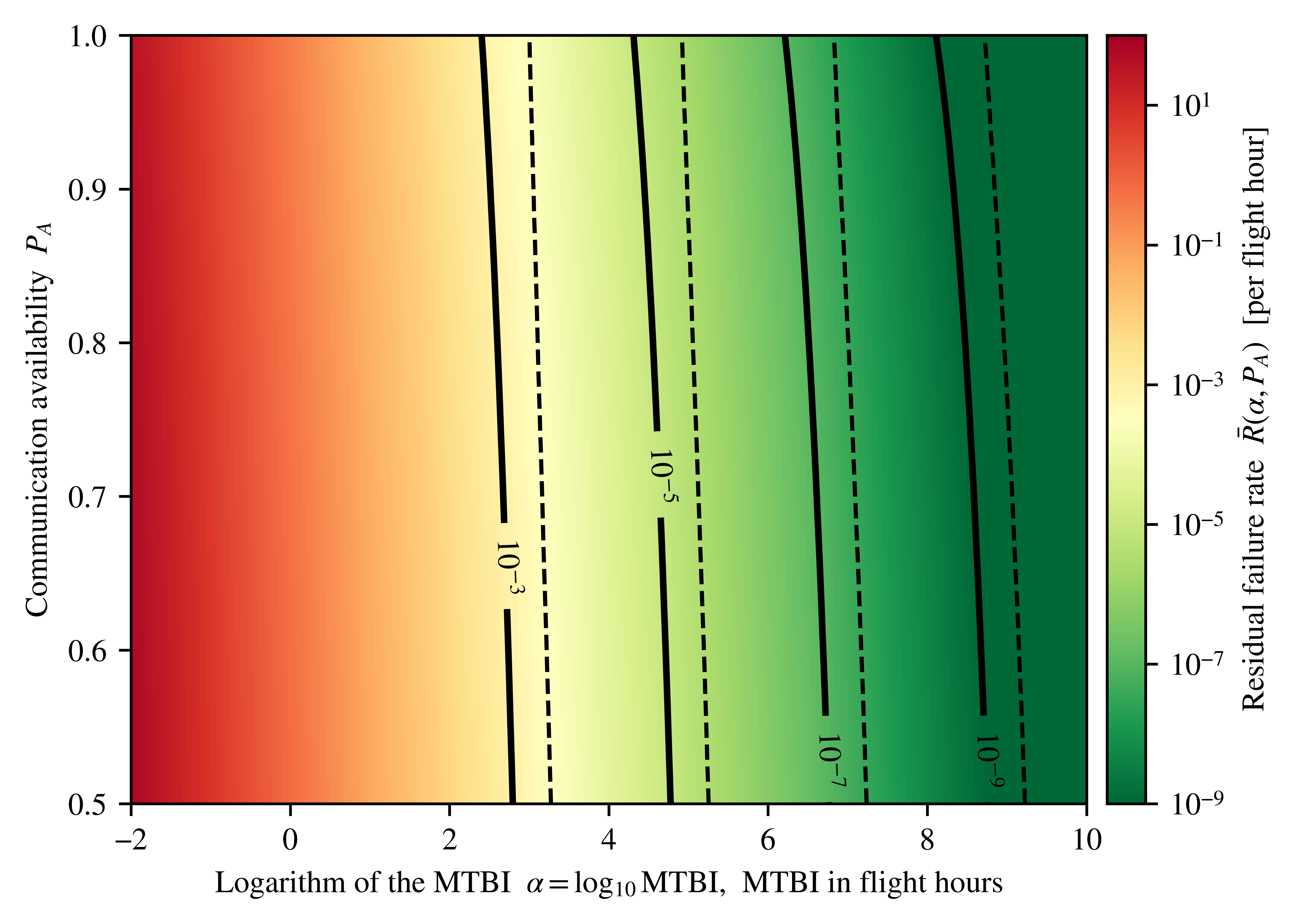}
\caption{Risk map over $\alpha$ and $P_A$ using $40{,}000$ capability draws, $m_{\mathcal{U}}=60$~s, and $\varepsilon=0.25$~s. Color shows the mean residual failure rate per flight hour. Solid contours mark mean compliance with safety budgets of $10^{-3}$, $10^{-5}$, $10^{-7}$, and $10^{-9}$ per flight hour. Dashed contours use the 95th percentile of conditional rates. Their displacement records the additional requirement associated with parameter uncertainty.}
\label{fig: auto_avail}
\end{figure}

For the routine function, Figure~\ref{fig: auto_avail} shows that a higher MTBI provides the main reduction in the residual failure rate. It uses $m_{\mathcal{U}}=60$~s and $\varepsilon=0.25$~s. Across most of the plane, increasing the availability changes the rate by less than one severity class, which is why the constant risk contours are nearly vertical. The short latency leaves the complexity as the larger timing constraint.

For this same function, the assurance contours require about $0.5$ to $0.6$ more units of $\alpha$ than the mean contours. The exponential distribution of the frequency alone contributes $\log_{10}(-\ln\delta)\approx0.48$. Uncertainty in the urgency and the complexity adds the remainder. Thus, a design meeting the budget in expectation can fail the assurance criterion under the same parameter distribution.

Interpreting these contours requires checking the uncongested intervention assumption. The corner at $\alpha=-2$ has nominal utilization near $0.8$, outside that regime. The contour interpretation should therefore be restricted to allocations with low utilization. Nominal utilization remains below $10^{-4}$ along the plotted budget contours. At sufficiently large $\alpha$, the low intervention frequency meets each illustrated safety budget even without remote intervention.

\begin{figure}[htbp]
\centering
\includegraphics{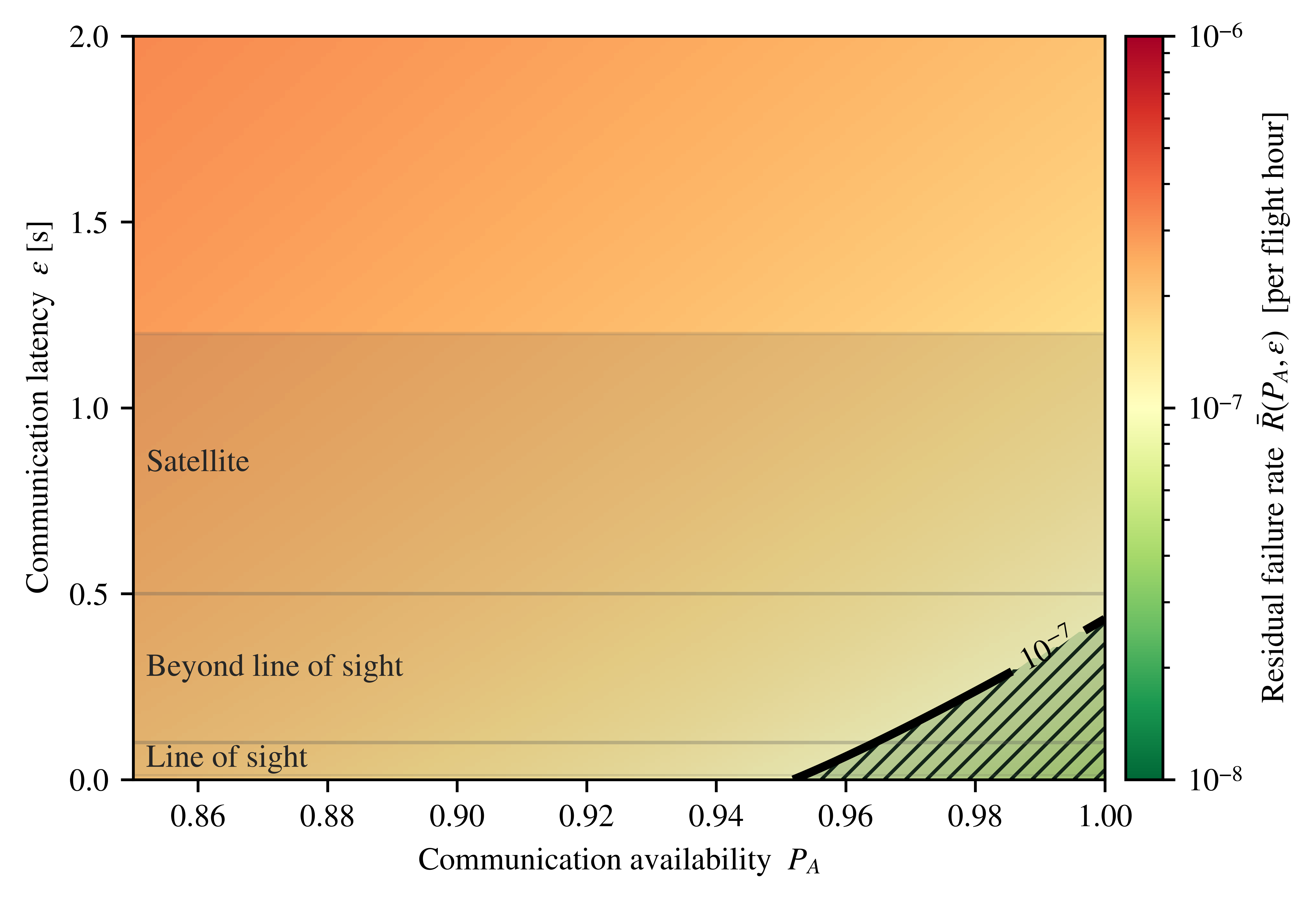}
\caption{Communication risk map with $m_{\mathcal{F}}=10^{-6}$ per flight hour, $m_{\mathcal{U}}=30$~s, and $m_{\mathcal{C}}=0.5$~s. Color shows the mean residual failure rate per flight hour. The solid boundary and hatched region indicate mean compliance with $10^{-7}$ per flight hour. No plotted allocation meets that budget at 95\% assurance under the assumed uncertainty. Horizontal bands show representative latencies for line of sight, beyond line of sight, and satellite links.}
\label{fig: comm_latency}
\end{figure}

For the second function, the performance of the communication subsystem determines whether mean compliance is reached. Figure~\ref{fig: comm_latency} fixes $m_{\mathcal{F}}=10^{-6}$ per flight hour, $m_{\mathcal{U}}=30$~s, and $m_{\mathcal{C}}=0.5$~s. This is an operating point at $\alpha=6$ with a nominal MTBI of $10^6$ flight hours. Its mean complexity is specified directly for this function and is shorter than the trend of the first chart would give. The safety budget is the hazardous value of $10^{-7}$ per flight hour. Under mean compliance, the tolerable latency rises from zero near $P_A=0.95$ to about $0.45$~s at unit availability. The boundary therefore translates the specified function and safety budget into a requirement on the communication subsystem.

The mean compliant region in this chart does not extend to 95\% assurance under the same uncertainty assumptions. The quantile remains above the safety budget even for a perfect link. Under the assumed capability distribution, improving the communication subsystem alone cannot establish this stronger compliance statement. A higher MTBI, a longer urgency, or a lower complexity is required.

This assurance failure and the tight mean latency limit reflect the effect of short sampled urgencies. Holding the triple at its means moves the unit availability latency boundary to about $2.7$~s. A gamma capability distribution with shape four and the same means moves it to about $1.9$~s. The $0.45$~s boundary is therefore conditional on the diffuse exponential capability distribution. The representative latency bands in the figure permit some line of sight and faster beyond line of sight links under mean compliance. The satellite band exceeds the largest tolerable latency in that illustration.

The sensitivity to short urgencies leaves little failure probability allowance for the latency. For fixed parameters, the allowed probability of an unsuccessful intervention is $R_{\mathrm{budget}}\mathrm{MTBI}$. When unavailability and complexity consume most of this allowance, a small added latency can determine compliance. In Figure~\ref{fig: comm_latency}, the nominal allowance is $0.1$. Under the exponential capability distribution, the complexity alone consumes about half of it. A latency of about 1.5\% of the mean urgency consumes the remainder.

The different constraints in these two cases show why the requirement on the communication subsystem depends on the function being supported. A better link improves the level-of-safety only for interventions that can still be completed in time. Stringent safety budgets also require a sufficiently high MTBI, even when the criticality ratio is small. The result therefore concerns a specified function and operating setting. It does not assign one safety capability to the aircraft through an ordinal autonomy label.

\section{Evaluation Procedure with a Closed Loop Simulator\label{sec: evaluation}}
\begin{figure}[!tp]
\centering
\definecolor{figdata}{HTML}{56B4E9}
\begingroup
\hypersetup{linkcolor=black}
\begin{tikzpicture}[
    font=\footnotesize,
    line width=0.5pt,
    node distance=3.5mm and 5mm,
    >={Latex[length=2mm, width=1.6mm]},
    every node/.style={inner xsep=2mm, inner ysep=1mm},
    io/.style={draw, trapezium, trapezium left angle=72, trapezium right angle=108,
        trapezium stretches=true, minimum height=10mm, text width=30mm,
        align=center, fill=figdata!25},
    proc/.style={draw, minimum height=10mm, text width=78mm, align=center,
        fill=black!4},
    data/.style={draw, trapezium, trapezium left angle=72, trapezium right angle=108,
        trapezium stretches=true, minimum height=10mm, text width=62mm,
        align=center, fill=figdata!25},
    test/.style={draw, diamond, aspect=2.8, text width=32mm, align=center,
        inner sep=0pt, fill=white},
    frame/.style={draw=black!60, dashed, line width=0.4pt, inner sep=2.5mm},
    framelabel/.style={fill=white, inner sep=1.5pt, text=black!75},
    lbl/.style={fill=white, inner sep=1.5pt, align=center}
]
\node[proc] (s1) {1.\ Reference mission for the specified\\function and operating setting};
\node[proc, below=of s1] (s2) {2.\ Intervention triggers from the alert set\\and failure boundary from the safe set};
\node[proc, below=of s2] (s3) {3.\ Sweep of the alert thresholds\\and the onboard competence};
\node[proc, below=of s3] (s4) {4.\ Closed loop simulation and censored maximum\\likelihood estimate of $(\mathrm{MTBI},\mathcal{U},\mathcal{C})$};
\node[proc, below=of s4] (s5) {5.\ Fit of $m_{\mathcal{C}}(\alpha)$ and $m_{\mathcal{U}}$ over the\\configurations of one alert threshold setting};
\node[data, below=of s5] (pia) {Capability distribution $\pi_A(\cdot\mid\alpha)$ with\\$\alpha$ the logarithm of the estimated MTBI};
\node[proc, below=9mm of pia] (risk) {Monte Carlo risk evaluation, Eqs.~\eqref{eq:mc_estimator} and~\eqref{eq:safety_budget},\\giving the mean $\bar{R}$ and the $(1-\delta)$ quantile};
\node[test, below=of risk] (dec) {Within $R_{\mathrm{budget}}$ at\\assurance $1-\delta$?};
\node[data, below=6mm of dec] (out) {Acceptable allocation $(\alpha,\theta_C)$\\and risk maps};
\node[io, left=of s1] (task) {Mission definition};
\node[io, left=of s4] (plant) {Vehicle dynamics\\model};
\node[io, left=of risk] (chan) {Channel model\\$\theta_C=(P_A,\varepsilon)$};
\begin{scope}[on background layer]
\node[frame, fit=(s1) (s2) (s3) (s4) (s5) (pia)] (frA) {};
\node[frame, fit=(risk) (dec) (out)] (frB) {};
\end{scope}
\node[framelabel, anchor=west] at ($(frA.north west)+(3mm,0)$) {Estimation of the autonomy triple, link idealized};
\node[framelabel, anchor=west] at ($(frB.north west)+(3mm,0)$) {Risk evaluation};
\draw[->] (task) -- (s1);
\draw[->] (plant) -- (s4);
\draw[->] (chan) -- (risk);
\draw[->] (s1) -- (s2);
\draw[->] (s2) -- (s3);
\draw[->] (s3) -- (s4);
\draw[->] (s4) -- (s5);
\draw[->] (s5) -- (pia);
\draw[->] (pia) -- (risk);
\draw[->] (risk) -- (dec);
\draw[->] (dec) -- node[lbl, right=1pt, pos=0.4] {yes} (out);
\coordinate (fbA) at ($(frA.east)+(13mm,0)$);
\draw[->] (dec.east) -- node[lbl, above=1pt, pos=0.3] {no} (dec.east -| fbA) -- (s3.east -| fbA) -- (s3.east);
\path (dec.east -| fbA) -- (s3.east -| fbA) node[midway, lbl] {revise the\\autonomy\\subsystem};
\draw[->] (dec.west) -- node[lbl, above=1pt, pos=0.25] {no} (dec.west -| chan) -- (chan.south);
\path (dec.west -| chan) -- (chan.south) node[midway, lbl] {upgrade the link};
\end{tikzpicture}
\endgroup
\caption{Proposed parameter estimation and risk evaluation procedure. Simulations with an idealized link supply the capability distribution of the autonomy parameters. Candidate communication parameters enter the subsequent risk calculation. An allocation that does not meet the safety budget at the chosen assurance level prompts revision of the autonomy subsystem or the communication subsystem.}
\label{fig: eval_flowchart}
\end{figure}
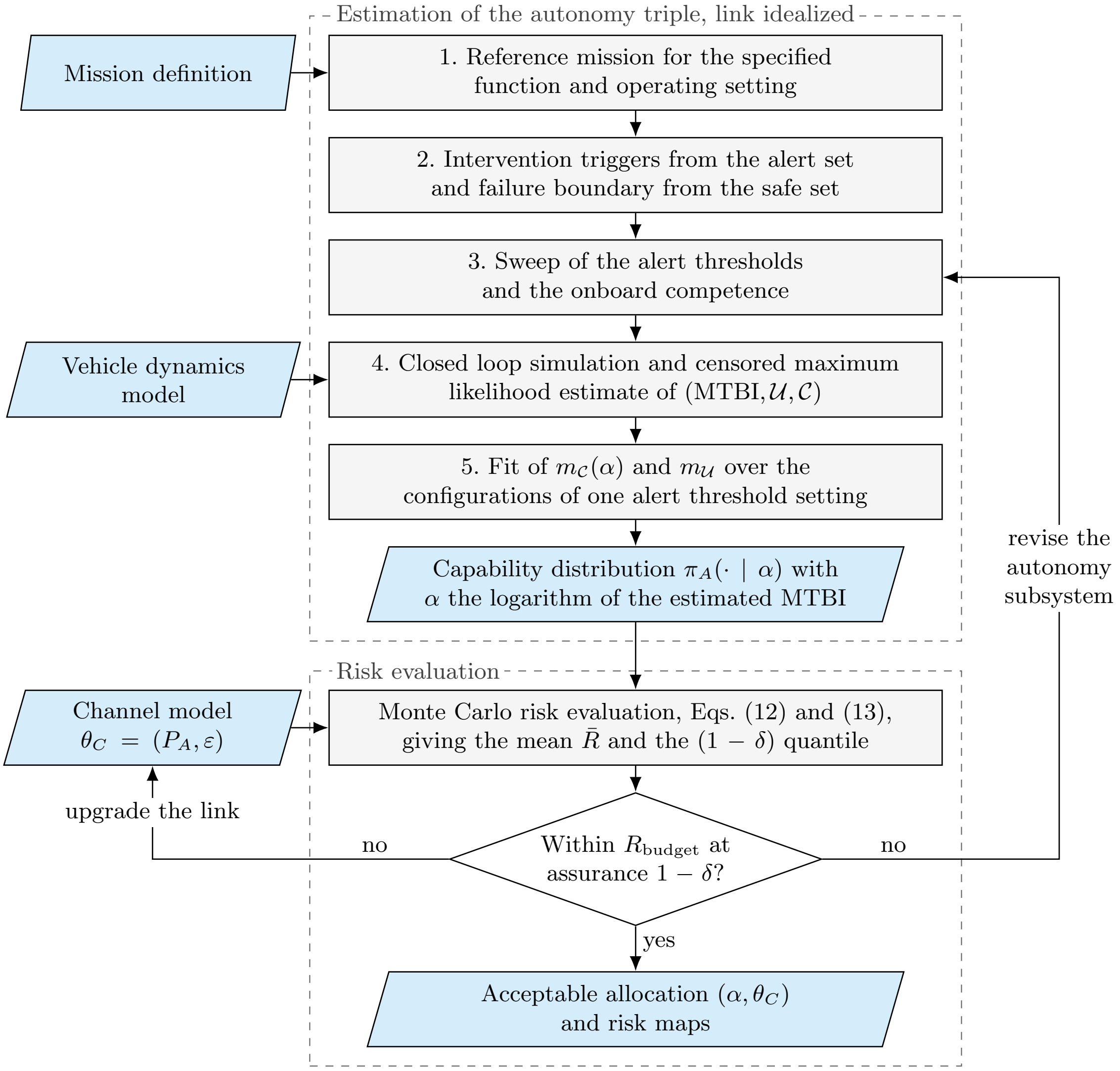

Applying these function specific requirements to an actual design requires measured autonomy parameters. We now describe a proposed procedure for estimating them from a closed loop simulator. Figure~\ref{fig: eval_flowchart} separates the estimation of the autonomy parameters with an idealized link from risk evaluation with a candidate communication subsystem. The procedure requires a vehicle model, a specified function and operating setting, intervention triggers, and a failure boundary.

Measuring those parameters first requires a mission definition and explicit boundaries for the demand and for failure. An alert set defines when assistance is requested. A larger safe set defines the modeled failure boundary. Example triggers include slant range $d(t)\notin\mathcal{R}_{\mathrm{alert}}$, heading error $|e_\psi(t)|>\psi_{\mathrm{tol}}$, and control hold beyond $\tau_{\mathrm{hold}}$. The interval between interventions runs from renewed nominal operation to the next trigger. The intervention window $W$ begins at the trigger and ends at failure if the intervention does not succeed.

With those boundaries defined, the analyst varies onboard competence, contingency logic, and alert thresholds to measure their effects on the autonomy parameters. Each configuration produces its own MTBI, urgency, and complexity. Earlier alerts can raise the intervention frequency while leaving a longer urgency. Changes in onboard competence can also change which situations reach the operator. The measurements must therefore test the assumed relations among the MTBI, the urgency, and the complexity. The trend used in the illustration is one possible relation.

Each configuration in this sweep is simulated repeatedly with an idealized link to estimate the autonomy parameters. Under the exponential interval model, MTBI estimation includes exposure from runs that end without a demand. During a demand, the first observed event is either successful completion of the intervention or exit from the safe set. Under the independent exponential timing assumptions, the complexity and the urgency are each estimated as the total observed demand time divided by the corresponding event count. The two counts are completed interventions and exits from the safe set. These are the maximum likelihood estimates for censored exponential observations. Finite estimates require observations of the corresponding events.

Using an idealized link for these measurements keeps the estimated autonomy parameters separate from the performance of the communication subsystem. The estimated frequency locates a configuration on the $\alpha$ scale. Uncertainty in that estimate supplies the frequency component of the capability distribution. The candidate channel model enters only during joint risk evaluation.

The final step uses these estimates to replace the assumed parameter values and uncertainty distributions. A fitted relation between complexity and $\alpha$ replaces the assumed $m_{\mathcal{C}}(\alpha)$. Urgency estimates can be pooled only when they support a common urgency for the configurations being compared. Otherwise, the configurations require separate urgency parameters. Event counts and exposure also support narrower uncertainty distributions, such as a gamma distribution for the frequency. These measurements would test the assumptions of the illustration before the resulting risk maps are used to allocate requirements.

\section{Conclusions\label{sec: conclusion}}
We developed an analytical model for the level-of-safety of partially autonomous machine systems as a function of the joint performance of their autonomy and communication subsystems. The level-of-safety is a residual failure rate that combines the intervention frequency with the probability that a remote intervention is unavailable or too late. The criticality ratio prescribes the shape of the tradeoff between the MTBI and the availability at a constant level-of-safety. For a perfect link at a ratio of 0.01, the MTBI needed to meet a given safety budget is two orders of magnitude lower than without remote intervention. At a ratio of one, even a perfect link only halves the required MTBI. These statements hold for fixed autonomy parameters. A positive latency reduces the benefit further at the same ratio, within the bounds of Eq.~\eqref{eq:rho_bounds}.

This tradeoff is stated on a continuous scale of the MTBI. The logarithm of the MTBI measures how often a specified function requires human intervention in a given operating setting, and it can be estimated from simulated or operational data. The urgency and the complexity complete the description of the autonomy subsystem and enter the tradeoff through the ratio. The constant risk frontiers then show how much a better link can lower the required MTBI for each value of the ratio. With exponentially distributed intervention window and nominal completion time, the model is in closed form.

When the uncertainty in the autonomy parameters enters that model, the limiting subsystem and the assurance result depend on the function. A higher MTBI has the main effect in the routine case. For the rare function with a short urgency, mean compliance limits the latency to about $0.45$~s at unit availability. No link meets the hazardous safety budget at 95\% assurance under the assumed exponential capability distribution. Narrower distributions relax the mean latency limit. Thus, the numerical requirement on the communication subsystem depends on the uncertainty assigned to the function.

These numerical requirements remain conditional on independent demands and low operator utilization. The model omits queuing and human errors that do not depend on the length of the intervention window. The exponential intervention window can overstate latency effects when alerting guarantees minimum lead time. The base availability model also treats an outage as lasting throughout the intervention window. The finite outage extension illustrates the sensitivity to that choice, while interruptions during an exchange remain unresolved.

Testing these assumptions for a candidate design requires the estimates of the autonomy parameters specified by the proposed closed loop simulator procedure. The procedure has not been demonstrated here, and the illustration does not establish empirical safety performance for an aircraft. Application requires evidence for the capability distribution and an appropriate allocation of the safety budget. Other partially autonomous machine systems, such as road vehicles, ships, and spacecraft, would also require suitable failure definitions, exposure units, and safety budgets.

\section*{Acknowledgment}
This work was supported by the National Aeronautics and Space Administration (NASA) University Leadership Initiative (ULI) program under project \textit{Autonomous Aerial Cargo Operations at Scale}, via grant No. 80NSSC21M071 to the University of Texas at Austin. Any opinions, findings, conclusions, or recommendations expressed in this material are those of the authors and do not necessarily reflect the views of the project sponsor.

\bibliography{refs}

\end{document}